# Stabilization of dipole solitons in nonlocal nonlinear media

Fangwei Ye, Yaroslav V. Kartashov, and Lluis Torner

*ICFO-Institut de Ciencies Fotoniques, and Universitat Politecnica de Catalunya, Mediterranean Technology Park, 08860 Castelldefels (Barcelona), Spain*

We address the stabilization of dipole solitons in nonlocal nonlinear materials by two different approaches. First, we study the properties of such solitons in thermal nonlinear media, where the refractive index landscapes induced by laser beams strongly depend on the boundary conditions and on the sample geometry. We show how the sample geometry impacts the stability of higher-order solitons in thermal nonlinear media and reveal that dipole solitons can be made dynamically stable in rectangular geometries in contrast to their counterparts in thermal samples with square cross-section. Second, we discuss the impact of the saturation of the nonlocal nonlinear response on the properties of multipole solitons. We find that the saturable response also stabilizes dipole solitons even in symmetric geometries, provided that the input power exceeds a critical value.

PACS numbers: 42.65.Tg, 42.65.Jx, 42.65.Wi.

## 1. Introduction

Solitons in nonlinear media exist in a variety of forms and shapes. In addition to the nodeless amplitude distributions featured by fundamental solitons, there exist stationary states with complex spatial shapes, including multiple light spots. However, in uniform media with local response, such complex patterns tend to be dynamically unstable, except in the form of vector solitons under suitable conditions. The situation changes dramatically in the nonlocal nonlinear media, where in sharp contrast to the case of local media, out-of-phase bright solitons can attract each other [1] and may even form scalar bound states, in both one- [2-4] and two-dimensional [5-12] geometries. In addition to stationary multipoles [5,8], two-dimensional settings support even rotating structures [7,13]. Nonlocal media with different types of response have been used to demonstrate steady soliton spiraling [14,15] mediated by long-range interaction forces between the light spots.

Stationary two-dimensional multipole-mode solitons have been observed in thermal nonlinear media [9]. Such materials, where nonlocality appears as a result of heat transport inside the sample, may exhibit a particularly strong nonlocal response [16-18]. Nevertheless, all stationary, nonrotating multipoles encountered so far experimentally in nonlocal materials, such as nematic liquid crystals or thermal media, are weakly unstable, with the exception in liquid crystals where dipoles were found to be stable in a very narrow region at low powers [19]. A similar picture holds for different nonlocal materials [8,10,11]. Notice, that the possibility of stabilization of multipoles by setting them into rotation was suggested theoretically [7].

An important aspect associated with nonlocal nonlinear materials is that the material boundaries often play a central role in the propagation of light. This is the case, for example, in liquid crystals where boundaries provide anchoring of molecules and thus determine the nonlocality degree of the nonlinear response [1], and in thermal materials where steady-state refractive index distributions depend on the sample geometry and on the temperature distribution at its boundaries [17]. Therefore, one may expect that in materials with large range of nonlocality, boundaries may dramatically affect the stability of nonlinear excitations. The interaction with far-away boundaries is one of the potential mechanisms of stabilization of multipole solitons that we consider in this work.

On the other hand, most of the aforementioned papers where multipole solitons are addressed concentrate on the simplest model of Kerr nonlinear nonlocal response (where the nonlinear correction to the refractive index is determined by the first power of the light intensity) and do not take into account a potential saturation of the nonlocal nonlinearity. Such Kerr nonlocal response appears, e.g., in lead glasses featuring a thermal nonlinearity [14,16] and in nematic liquid crystals with reorientational nonlinearity [1,15]. In such media, a laser beam with intensity $I$ induces a refractive index profile described by the diffusion-type equation $\Delta \delta n = -I$ in thermal media or $\delta n - d\Delta \delta n = I$ in liquid crystals. A different phenomenological model was introduced in Ref. [20] where a cubic-quintic nonlocal nonlinearity was considered with a nonlinear contribution to the refractive index depending also the quadratic power of the light intensity. Such a response can be encountered, e.g., in atomic vapors [21,22].

In the first part of this paper we address the impact of the sample geometry on the stability of higher-order solitons in thermal nonlinear media. We report, for the first time to our knowledge, an example of the sample-geometry-mediated stabilization of complex soliton states. Thus, in contrast to perfectly square geometries, dipole solitons might be stable in samples with a rectangular cross-section. Such stabilization strongly depends on the soliton orientation inside the sample and on the sample geometrical aspect-ratio. We also find that only dipoles, but not higher-order solitons, can be made stable in such samples.

In the second part of this paper we study the role of a saturable nonlocal nonlinearity in the stabilization of dipole solitons that are unstable in samples made of liquid crystals and lead glasses with a square cross-section. We show that saturation results in stabilization of dipoles above a certain energy flow threshold, while all higher-order multipoles in such media were found to be unstable.

## 2. Geometry-mediated stabilization

First, we consider the propagation of a laser beam in thermal nonlinear media, described by the nonlinear Schrödinger equation for the slowly varying light field amplitude $A$ coupled to the steady-state heat transfer equation. Light beams propagating in thermal medium experience slight absorption and thus act as heat sources. The heat diffusion results in transversely inhomogeneous temperature distributions which depend on the temperatures at the sample boundaries and on the sample geometry. Temperature redistribution leads to a variation of local refractive index proportional to the temperature change [9,16]. The governing system of equations in dimensionless form thus writes as

$$i\frac{\partial q}{\partial \xi} = -\frac{1}{2}\left(\frac{\partial^2 q}{\partial \eta^2} + \frac{\partial^2 q}{\partial \zeta^2}\right) - nq, \tag{1a}$$

$$\frac{\partial^2 n}{\partial \eta^2} + \frac{\partial^2 n}{\partial \zeta^2} + \nu \frac{\partial^2 n}{\partial \xi^2} = -|q|^2. \tag{1b}$$

Here $q = (k_0^2 r_0^4 \alpha\beta / \kappa n_0)^{1/2} A$ is the dimensionless light field amplitude; $n = k_0^2 r_0^2 \delta n / n_0$ is proportional to the nonlinear change $\delta n$ in the refractive index $n_0$; $\alpha, \beta, \kappa$ are the optical absorption, thermo-optic, and thermal conductivity coefficients, respectively; the transverse $\eta$, $\zeta$ and longi-tudinal $\xi$ coordinates are scaled to the characteristic beam width $r_0$ and diffraction length $L_{\text{dif}} = k_0 r_0^2$, respectively.

The derivative $\partial^2 n / \partial \xi^2$ appearing in Eq. (1b) stands for heat transfer and nonlocality in the longitudinal direction, and the coefficient $\nu = (r_0 / L_{\text{dif}})^2$ characterizes the relative weight of such longitudinal effects. Since Eq. (1b) has to be solved together with the boundary conditions at all crystal boundaries, the analysis of the full system (1) with high accuracy requires significant computer resources. However, in typical experiments where the crystal length usually greatly exceeds the sample transverse dimensions and where laser beams are used with width $r_0 \gg \lambda$, one has $\nu \ll 1$ by many orders of magnitude, and it is justified to simplify the system (1) by neglecting nonlocality in the longitudinal direction. Under such conditions, one gets a truncated system that we employ in this study. Namely,

$$i\frac{\partial q}{\partial \xi} = -\frac{1}{2}\left(\frac{\partial^2 q}{\partial \eta^2} + \frac{\partial^2 q}{\partial \zeta^2}\right) - nq,$$
$$\frac{\partial^2 n}{\partial \eta^2} + \frac{\partial^2 n}{\partial \zeta^2} = -|q|^2.$$
(2)

To illustrate that such a truncation is fully justified in the usual suitable experiments, we performed several estimates and calculations. For example, we performed an accurate comparison of the refractive index distributions obtained using both, the truncated and the complete models for different values of the parameter $\nu$ and for different input conditions. To keep the numerical resources needed within our availability, for simplicity we considered the propagation of circularly symmetric input beams in cylindrical thermal samples. Note that the longitudinal derivatives have similar weight in different sample geometries.

At the first step, the dynamics of the input beam using the model (2) was calculated and the corresponding three-dimensional intensity $|q(\eta,\zeta,\xi)|^2$ and refractive index $n(\eta,\zeta,\xi)$ distributions were recorded. Then, the intensity distribution obtained using (2) was substituted into right-hand side of Eq. (1b). The resulting fully three-dimensional equation was solved for a given $|q(\eta,\zeta,\xi)|^2$ under the assumption that the input and output crystal facets are thermally insulating, to obtain the refractive index distribution resulting from the complete model (1). The maximal nonlinear contributions $n_{max}$ to the refractive index obtained with both models are compared in Fig. 1 for $\nu = 0.001$. The point is the difference between both cases is barely visible, despite the complex dynamics of the refractive index together with the substantial oscillations of $n_{max}$, and despite the huge value of $\nu$ used, a value several orders of magnitude larger than the values encountered in the experiments that we model in here. Note that for typical experimental parameters $r_0 \sim 25\ \mu$m, $L_{dif} \sim 1.5$ cm one gets $\nu \sim 3\times 10^{-6}$, almost three orders of magnitude smaller. This means that for sufficiently small values of $\nu$ it is completely justified to neglect the nonlocality in longitudinal direction. Notice also that this assumption was used in a number of previous experimental investigations where an excellent agreement between experimental observations and theoretical predictions was found [17,18,23]. We thus concentrate on the model (2).

In thermal media the conditions imposed at the boundaries of the sample greatly affect the entire optically induced refractive index profile. We are interested in situations where the refractive index distribution exhibits a geometry-induced anisotropy. This can be achieved by keeping the opposite crystal faces at different temperatures by means of external heat sinks [17], by putting sample faces in contact with materials having different thermal conductivities, or by fabricating samples with a rectangular cross-sections strongly elongated in one direction. Here we con-sider the latter case and assume that boundaries are kept at equal temperatures, i.e. we solve Eq. (2) with the boundary conditions $n,q|_{\eta \to \pm L_\eta/2}=0$, $n,q|_{\zeta \to \pm L_\zeta/2}=0$, where $L_\eta$ and $L_\zeta$ are sample dimensions along $\eta$ and $\zeta$ axes, respectively. Note that, formally, the refractive index at the boundary can always be set to zero because adding the constant background in the refractive index is equivalent to introducing a shift of soliton propagation constant. We set $L_\eta=40$, which corresponds to real experimental conditions, and vary $L_\zeta \leq L_\eta$. Finally, for further use, notice that Eqs. (2) conserve the energy flow $U$ and the Hamiltonian $H$

$$U = \iint_{-\infty}^{\infty} |q|^2 \, d\eta d\zeta,$$
$$H = \iint_{-\infty}^{\infty} \left[ \frac{1}{2}\left( \left|\frac{\partial q}{\partial \eta}\right|^2 + \left|\frac{\partial q}{\partial \zeta}\right|^2 \right) - \frac{1}{2}|q|^2 \iint_{-\infty}^{\infty} G(\eta,\zeta;\lambda,\mu)|q|^2 \, d\lambda d\mu \right] d\eta d\zeta, \quad (3)$$

where

$$G(\eta,\zeta;\lambda,\mu) = \sum_{l,m=1}^{\infty} \frac{4\sin(\pi l\eta/L_\eta)\sin(\pi m\zeta/L_\zeta)\sin(\pi l\lambda/L_\eta)\sin(\pi m\mu/L_\zeta)}{L_\eta L_\zeta [(\pi l/L_\eta)^2 + (\pi m/L_\zeta)^2]} \quad (4)$$

is the response function of a thermal medium with a rectangular cross-section.

We search for soliton solutions of Eq. (2) in the form $q(\eta,\zeta,\xi) = w(\eta,\zeta)\exp(ib\xi)$, here $w$ is a real function and $b$ is the propagation constant. The resulting two-dimensional differential nonlinear eigenvalue problems were solved with a standard relaxation method, starting with an appropriate initial guess of the mode profile. While for fundamental solitons $w(\eta,\zeta)$ has no nodes, the dipole solitons are characterized by two bright spots with a $\pi$ phase jump between them. We found two different types of dipoles in rectangular samples: the nodal line separating two spots in one of them is perpendicular to the longest side of the sample [Fig. 2(a)], while in the other soliton it is parallel to the longest sample side [Fig. 2(b)]. In both cases the width of the refractive index distribution far exceeds the widths of individual spots in the soliton profile [Figs. 2(c) and 2(d)]. A noteworthy feature is that while the symmetry of the refractive index profile in the central part of the sample reflects the dipole orientation, the overall refractive index distribution is still elongated in the $\eta$ direction since $L_\eta > L_\zeta$. This difference in the internal refractive index distributions gives rise to crucial differences in the stability properties of solitons, despite the fact that for equal values of propagation constants $b$ the functional shapes of the individual spots in dipole intensity distribution remain close for two different orientations. Still, for fixed $b$ the dipole oriented along the short $\zeta$ axis carries a higher energy flow than its counterpart oriented along the long $\eta$ axis.

Figures 3(a)-3(d) summarize the properties of dipoles whose nodal lines are perpendicular to the longest side of the sample [Fig. 2(a)]. For all values of $L_\zeta$ the energy flow is an increasing function of propagation constant. The larger the difference between lengths $L_\zeta$ and $L_\eta$ the larger the difference between energy flows of solitons corresponding to the same propagation constants. This difference becomes more pronounced for large $b$. The integral radii $R$ of individual spots forming dipole substantially decrease with $b$ [Fig. 3(b)], hence even strongly localized dipoles are affected by the sample geometry. The picture remains qualitatively similar for different dipole orientations. For both types of solitons the energy flow vanishes when $b \to 0$.

To gain an intuitive understanding of the stability properties of the dipoles, we study their Hamiltonian [24]. Figure 3(e) shows the Hamiltonian versus the energy flow for both types of dipoles in a rectangular sample. The dependence $H(U)$ for fundamental solitons in the same set-ting is also shown for the sake of comparison. Dipole solitons feature higher values of Hamiltonian than fundamental solitons with the same energy flow, in agreement to the higher-order nature of the former. The important result revealed by the plot is that the dipoles having different orientations carry slightly different Hamiltonians: Dipoles oriented along the $\eta$ axis exhibit a lower Hamiltonian than their counterparts oriented along the $\zeta$ axis [Fig 3(f)]. Since dipoles are excited states, the comparison of Hamiltonians does not prove the stability of solution with lower $H$, but such a comparison does suggest that that dipoles oriented along the $\eta$ axis may be more stable than their counterparts oriented along $\zeta$ axis [24]. Such expectation is fully consistent with the finding of our rigorous numerical simulations, described below. Notice that examples of stability of multiple solutions for a given Hamiltonian in other similar physical settings are known (see, e.g., [25] for an example in the case of quadratic solitons).

To elucidate the actual impact of sample geometry on the soliton stability we performed comprehensive, rigorous numerical simulations of the governing system of equations (2) with the input conditions $q|_{\xi=0}=w(1+\rho)$, where $\rho$ stands for the broadband input noise with the variance $\sigma_{\text{noise}}^2=0.01$. The numerical method we used to solve Eq. (2) is a symmetrized split-step Fourier scheme, with the linear and nonlinear parts treated alternately in Fourier and physical space. The central finding of this paper is that dipole solitons that are unstable in thermal media with square cross-section [9] can become stable in samples with rectangular cross-section for the proper orientation of dipole with respect to the sample. This is the case for dipoles whose nodal lines are perpendicular to the longer side of the sample. Figure 4(a) shows the stable propagation of such a dipole in the presence of large input perturbations. In clear contrast, dipoles oriented in the perpendicular direction are always unstable [Fig. 4(b)] and exhibit instabilities that might be much stronger than instabilities for dipoles having similar $b$ value in square $L_\eta \times L_\eta$ samples [Fig. 4(c)].

This constitutes the first example of the stabilization of complex soliton state mediated by the sample geometry in media with a long-range nonlocality of nonlinear response.

The domains of stability are depicted in Fig. 3(a) on the corresponding $U(b)$ curves. One can see that at $L_\zeta = 22$ the stability domain is approximately given by $b \in [0, 9.8]$, while at $L_\zeta = 15$ it expands up to $b \in [0, 23]$. Therefore, there exists a critical value $b_{cr}$ of propagation constant (or, equivalently critical energy flow $U_{cr}$) below which the dipoles become stable. The critical propagation constant $b_{cr}$ is a monotonically decreasing function of $L_\zeta$ [Fig. 3(c)]. The stability domain is wider in more elongated samples, and at $L_\zeta \to L_\eta$ the stability domain disappears. The critical energy flow also rapidly increases with decrease of $L_\zeta$ [Fig. 3(d)]. All instabilities encountered for the dipole solitons are oscillatory: unstable dipoles with $b > b_{cr}$ decay into fundamental solitons via progressively growing oscillations of amplitudes of bright spots. For dipoles whose nodal lines are parallel to the longer sample side the decay might be accompanied by the rotation of spots [Fig. 4(b)].

Note that the oscillatory instabilities of unstable dipoles may make them appear stable in experiments conducted with short samples. Thus, in typical settings (e.g., $\beta \sim 1.4 \times 10^{-5} \, \text{K}^{-1}$, $\alpha \sim 0.01 \, \text{cm}^{-1}$, $\kappa \sim 0.7 \, \text{W m}^{-1}\text{K}^{-1}$, $1 \times 1 \, \text{mm}^2$ sample cross section, and beam width $25 \, \mu\text{m}$ [16]), our simulations show that with input powers of a few Watts, the dipole instability manifest itself beyond $20 \, \text{cm}$, hence beyond sample length. Increasing the input power strongly enhances the instability strength, thus making it visible at shorter and shorter sample lengths.

Importantly, we found that the geometry-mediated stabilization holds only for dipole solitons. All higher-order solitons, including triple-mode solitons, quadrupoles, and necklaces were found to be unstable, even in strongly elongated samples with the ratio of dimensions $L_\eta / L_\zeta \sim 10$.

For example, Fig. 5 shows the decay of a triple-mode soliton whose nodal lines are perpendicular to the longer side of the sample for $b$ and $L_\zeta$ values that corresponds to geometry where a dipole soliton is stable. Still, this does not rule out the possibility that more complex sample geometries (rhomboidal, circular, hexagonal, etc) could lead to the stabilization of particular higher-order soliton states. We stress that linear absorption does not change our results. It gradually and slightly (on available crystal lengths) decreases the power of the beam upon propagation. Since dipoles are stable below an energy flow threshold, addition of slight absorption does not lead to destabilization of solitons that are otherwise stable.

### 3. Saturation-mediated stabilization

In this section, we address a second potential mechanism for stabilization of multipole solitons in nonlocal media. Specifically, we consider light propagation in two-state homogeneously broadened atomic vapor [21,22]. Such medium may exhibit a strong saturation as well as nonlocality. We further assume that the mean-free path of the atoms in the illuminated regions is much smaller than the width of the laser beam so that the atomic transport has a diffusive nature. Under such conditions the steady-state system of equations describing the laser-atom interaction can be written as [21]:

$$i\frac{\partial q}{\partial \xi} = -\frac{1}{2}\left(\frac{\partial^2 q}{\partial \eta^2} + \frac{\partial^2 q}{\partial \zeta^2}\right) - nq,$$
$$n - \frac{d}{1+S|q|^2}\left(\frac{\partial^2 n}{\partial \eta^2} + \frac{\partial^2 n}{\partial \zeta^2}\right) = \gamma \frac{|q|^2}{1+S|q|^2}.$$

(5)

Here $q=[\alpha(1-n_0)k_0^2r_0^2/\gamma_0 n_0]^{1/2}A$ is the dimensionless light field amplitude; $\alpha$ is the unsaturated resonant absorption coefficient; $\gamma_0$ is the atomic polarization decay rate; $k_0$ is the wavenumber; $n=k_0^2r_0^2\delta n/n_0$ is proportional to the nonlinear contribution $\delta n$ to the unperturbed refractive index $n_0$; as before, the transverse $\eta,\zeta$ and longitudinal $\xi$ coordinates are scaled to the characteristic beam width $r_0$ and diffraction length $L_{\text{dif}}=k_0r_0^2$, respectively; $d=D/\gamma_0 r_0^2$ is the unsaturated nonlocality degree with $D$ being the diffusion constant; the parameter $S=n_0/(1-n_0)k_0^2r_0^2$ characterizes the saturation degree; $\gamma=\pm 1$ stands for focusing (defocusing) nonlinearity and corresponds to settings when the laser is blue- (red-) detuned from the optical resonance frequency. Here we consider only focusing media and thus set $\gamma=+1$.

In the limiting case $S\to 0$ the system of Eqs. (5) transforms into a standard Kerr model of the nonlocal response with a single parameter $d$ characterizing the nonlocality degree. However, in the most general case with $S\neq 0$ the nonlocality degree depends on the spatial profile and peak intensity of the laser beam. In particular, the nonlocality degree decreases when $|q|^2\to\infty$. This model, introduced in Ref. [21], leads to a reasonable agreement with experimental observations of the collapse suppression of two-dimensional Gaussian-like laser beams. Recently a related, but different model describing beam propagation in vapors where nonlocality emerges from ballistic transport of the excited atoms was introduced in Ref. [26]. The role of saturation in the stabilization of higher-order two-dimensional solitons in vapors was not studied.

Typical field shapes and refractive index distributions for dipole solitons in saturable nonlocal media are shown in Fig. 6. The two maxima in the refractive index distribution are located approximately in the points where the field intensity reaches its maximal value. The local minimum appearing between the bright spots in the refractive index distribution is most pro-nounced for small $d$ values, but it washes out gradually when increasing the nonlocality degree (which is accompanied by an overall increase of the width of refractive index distribution notably exceeding the width of the intensity distributions). When decreasing the propagation constant, the individual spots in the dipole soliton profile broaden and the distance between spots increases [compare Figs. 6(a) and 6(b) corresponding to different $b$ values]. A similar scenario is encountered at very high energy flows, when the nonlinearity saturation weakens the effective nonlocality. Also, the forces acting between the out-of-phase constituents of a dipole soliton (hence, the width of the intensity distribution and the separation between spots in the soliton profile) are minimal at a certain intermediate value of the propagation constant. Note that the individual spots in the dipole profile that are elliptical at moderate energy flows become almost circularly symmetric at both, low and high energies.

The properties of dipole solutions in saturable media are summarized in Fig. 7. For all values of the $d$ and $S$ parameters the energy flow is a monotonically increasing function of the propagation constant as in the case of thermal nonlinearity. Increasing the nonlocality degree $d$ or the saturation parameter $S$ at a fixed $b$ causes an increase of the corresponding energy flow. The energy flow vanishes when $b \to 0$, but it diverges when $b$ approaches the upper cutoff for dipole existence, given by $1/S$. Besides dipole solitons we found a number of higher-order soliton solutions (not shown here) that can be arranged into lines or necklace-like configurations [9].

As in the case of thermal media we elucidated the impact of nonlinearity saturation on the soliton stability by conducting comprehensive numerical simulations of the evolution Eq. (5) with input conditions in the form of perturbed dipole solitons. The central finding is that saturation leads to stabilization of dipole solitons provided that their energy flows exceed certain critical value $U_{cr}$. This is in clear contrast to dipole solitons in the nonlocal Kerr media ($S=0$) where instability is enhanced with increasing energy flow. The domains of stability are depicted in Figs. 7(a) and 7(b) on the corresponding $U(b)$ curves. One can see that, for example, at $S = 0.06$ and $d = 10$ dipole solitons are stable at $U > U_{cr} \approx 174$ (or, equivalently, above the critical propagation constant value, $b_{cr} = 1.4$). Figure 8 shows examples of propagation of stable and unstable representatives of the dipole soliton family, with energy flows below [Fig. 8(a)] and well above [Fig. 8(b)] the critical one. In both cases, simulations were conducted in the presence of considerable input perturbations. While dipoles with energy flow below $U_{cr}$ tend to decay into fundamental solitons via progressively growing amplitude oscillations, their stable counterparts experience only a small reshaping and preserve their structure over huge distances.

We found that increasing the saturation degree results in a substantial expansion of the stability domain and in a reduction of the critical energy flow. Thus, at $d = 10$ the critical energy flow for $S = 0.02$ amounts to $U_{cr} \approx 354$, while for $S = 0.1$ one has $U_{cr} \approx 126$. The critical energy flow is a monotonically decreasing function of saturation parameter [Fig. 7(c)]. Smaller degrees of nonlocality also result in expansion of the stability domain. This is clearly visible in Fig. 7(a). Notice that $U_{cr}$ is quite small already at $d \sim 5$. Thus, for properly selected set of parameters $d$ and $S$, dipole solitons in our system can be stable almost in the entire existence domain. This is in clear contrast to the case of Kerr nonlocal media, where dipoles are unstable for any nonlocality degree. Importantly, as in the case of thermal nonlinear media, we found that saturation of nonlocal response can stabilize only dipoles, but not higher-order multipole solitons, including triple-mode ones, quadrupoles, and necklaces. All such higher-order states appear to be oscillatory unstable.

## 4. Conclusions

In conclusion, in this paper we addressed the possibility of stabilization of multipole-mode solitons in thermal and in saturable nonlocal media. We predicted the stabilization of dipole solitons in thermal media with a rectangular cross-section. To the best of our knowledge, this is the first example of geometry-mediated stabilization of soliton states in nonlocal materials with long-range nonlocality. Our conclusion is intended to hold for physical settings which can be accurately modeled by the system (2). We studied the particular case of thermal nonlinearity, but similar phenomena may take place in other nonlocal materials, such as liquid crystals, where boundaries strongly impact the dynamics of light beam propagation. We also studied the effect of saturation of nonlocal nonlinear response on stabilization of multipole solitons. We found that saturation can stabilize dipoles provided that their energy flow exceeds a critical value, but can not stabilize higher-order multipoles.

## 5. Acknowledgements

This work has been supported in part by the Government of Spain through the Ramon-y-Cajal program and through the grant TEC2005-07815.

**Figure captions**

Figure 1 (color online). Comparison of maximal nonlinear contributions to refractive index induced by the input beam $q|_{\xi=0} = A\exp(-r^2/W^2)\exp[-(r-4)^2/W^2]$ with $A=1$, $W=4$ in the cylindrical thermal sample with the radius $R=20$ when beam propagation is described by the truncated two-dimensional model (black line) and three-dimensional model (red line) at $\nu=0.001$. The difference between both cases is barely visible, despite the fact that the value $\nu=0.001$ is several orders of magnitude larger than those typically encountered in the experiments that we model in here. All quantities are plotted in arbitrary dimensionless units.

Figure 2 (color online). Field modulus (top row) and corresponding refractive index (bottom row) distributions for dipole solitons with two different orientations at $b=3$, $L_\eta=40$, $L_\zeta=22$. All quantities are plotted in arbitrary dimensionless units.

Figure 3 (color online). Energy flow (a) and integral radius of individual spots (b) for horizontally oriented dipole soliton versus propagation constant for different $L_\zeta$ values. Black curves correspond to stable branches, while red curves show unstable branches. Critical propagation constant (c) and energy flow (d) for stabilization of dipole solitons versus $L_\zeta$. (e) Hamiltonian versus energy flow for fundamental soliton (black line) and dipole solitons oriented along $\eta$ axis (red line) and $\zeta$ axis (green line) at $L_\zeta = 22$. (f) The difference of Hamiltonians $\delta H = H_\eta - H_\zeta$ as a function of energy flow $U$ at $L_\zeta = 22$. In all cases $L_\eta = 40$. All quantities are plotted in arbitrary dimensionless units.

Figure 4 (color online). (a) Stable propagation of horizontally oriented dipole and (b) decay of vertically oriented dipole at $b = 3$, $L_\eta = 40$, $L_\zeta = 22$. (c) Decay of dipole with $b = 3$ in square sample with $L_\eta = L_\zeta = 40$. Field modulus distributions at different distances are shown. All quantities are plotted in arbitrary dimensionless units.

Figure 5 (color online). Decay of horizontally oriented perturbed triple-mode soliton at $b = 3$, $L_\eta = 40$, $L_\zeta = 22$. Field modulus distributions at different distances are shown. All quantities are plotted in arbitrary dimensionless units.

Figure 6 (color online). Field modulus (left column) and refractive index (right column) distributions for dipole solitons corresponding to (a) $b = 0.8$ and (b) $b = 1.8$ at $S = 0.06$, $d = 10$. All quantities are plotted in arbitrary dimensionless units.

Figure 7 (color online). (a) Energy flow versus propagation constant for $d=5$ (curve 1), 10 (curve 2), and 20 (curve 3) at $S=0.1$. (b) Energy flow versus propagation constant for $S=0.02$ (curve 1), 0.06 (curve 2), and 0.1 (curve 3) at $d=10$. Black curves correspond to stable branches, red curves correspond to unstable branches. (c) Critical energy flow versus saturation parameter at $d=10$. All quantities are plotted in arbitrary dimensionless units.

Figure 8 (color online). Propagation dynamics of (a) unstable dipole with $b=0.8$ and (b) stable dipole with $b=1.8$ in the presence of broadband input noise with variance $\sigma_{noise}^2=0.01$. In both cases $d=10$, $S=0.06$. Field modulus distributions are shown at different distances. All quantities are plotted in arbitrary dimensionless units.

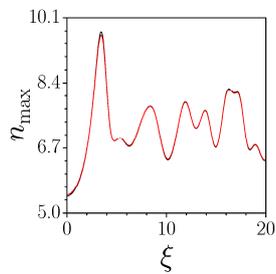

Figure 1 (color online).   Comparison of maximal nonlinear contributions to refractive index induced by the input beam $q|_{\xi=0} = A\exp(-r^2/W^2)\exp[-(r-4)^2/W^2]$ with $A=1$, $W=4$ in the cylindrical thermal sample with the radius $R=20$ when beam propagation is described by the truncated two-dimensional model (black line) and three-dimensional model (red line) at $\nu = 0.001$. The difference between both cases is barely visible, despite the fact that the value $\nu = 0.001$ is several orders of magnitude larger than those typically encountered in the experiments that we model in here. All quantities are plotted in arbitrary dimensionless units.

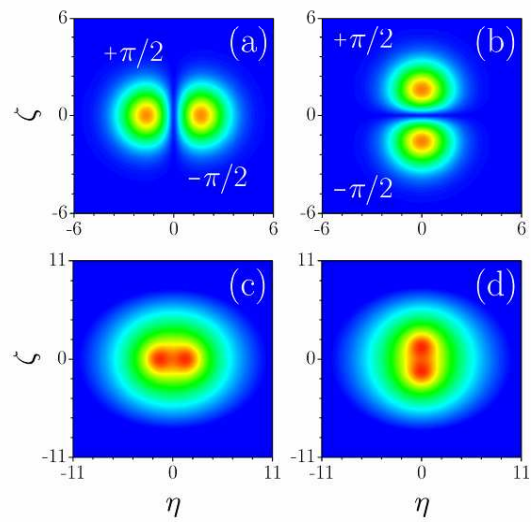

Figure 2 (color online). Field modulus (top row) and corresponding refractive index (bottom row) distributions for dipole solitons with two different orientations at $b = 3$, $L_\eta = 40$, $L_\zeta = 22$. All quantities are plotted in arbitrary dimensionless units.

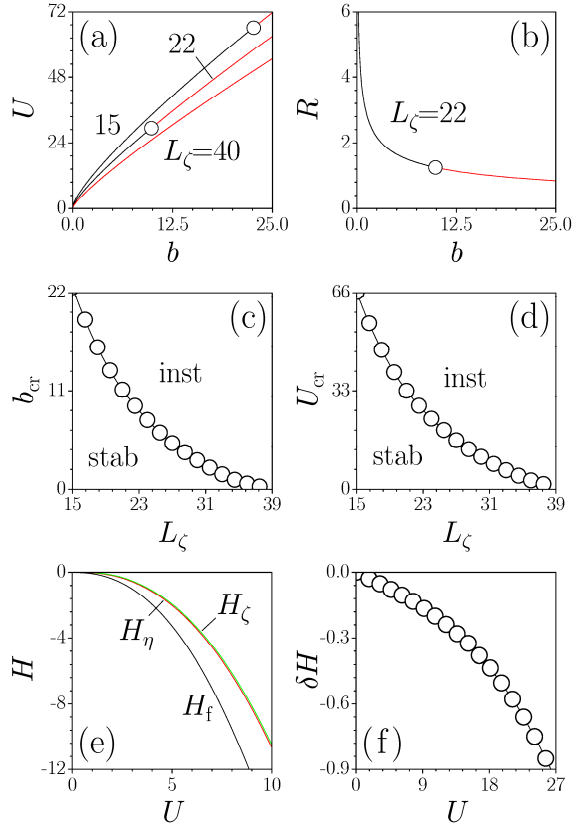

Figure 3 (color online). Energy flow (a) and integral radius of individual spots (b) for horizontally oriented dipole soliton versus propagation constant for different $L_\zeta$ values. Black curves correspond to stable branches, while red curves show unstable branches. Critical propagation constant (c) and energy flow (d) for stabilization of dipole solitons versus $L_\zeta$. (e) Hamiltonian versus energy flow for fundamental soliton (black line) and dipole solitons oriented along $\eta$ axis (red line) and $\zeta$ axis (green line) at $L_\zeta = 22$. (f) The difference of Hamiltonians $\delta H = H_\eta - H_\zeta$ as a function of energy flow $U$ at $L_\zeta = 22$. In all cases $L_\eta = 40$. All quantities are plotted in arbitrary dimensionless units.

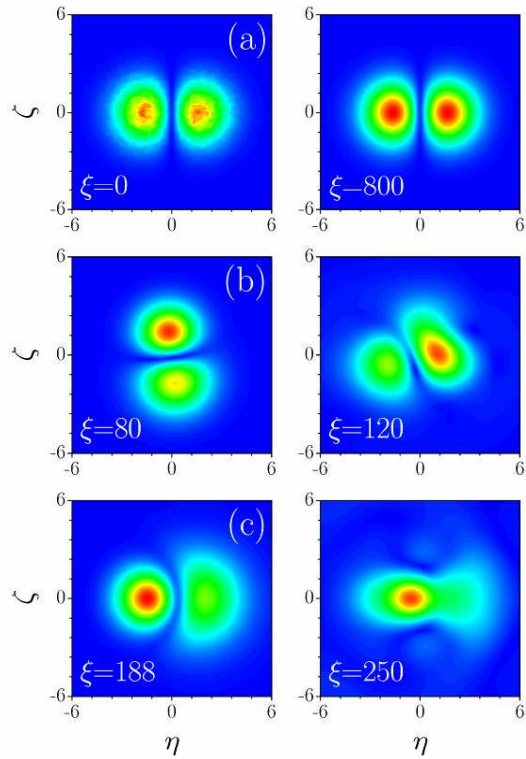

Figure 4 (color online). (a) Stable propagation of horizontally oriented dipole and (b) decay of vertically oriented dipole at $b=3$, $L_\eta = 40$, $L_\zeta = 22$. (c) Decay of dipole with $b=3$ in square sample with $L_\eta = L_\zeta = 40$. Field modulus distributions at different distances are shown. All quantities are plotted in arbitrary dimensionless units.

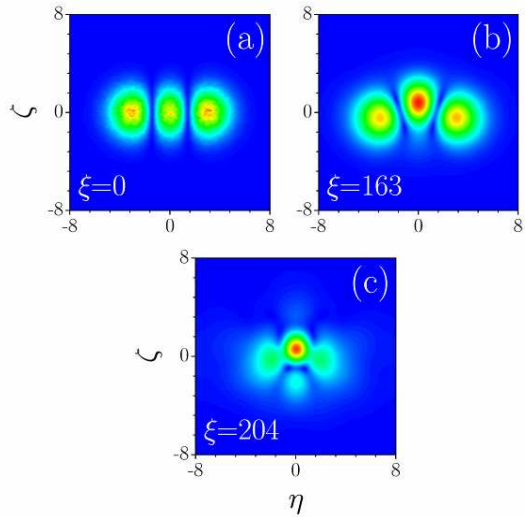

Figure 5 (color online). Decay of horizontally oriented perturbed triple-mode soliton at $b = 3$, $L_\eta = 40$, $L_\zeta = 22$. Field modulus distributions at different distances are shown. All quantities are plotted in arbitrary dimensionless units.

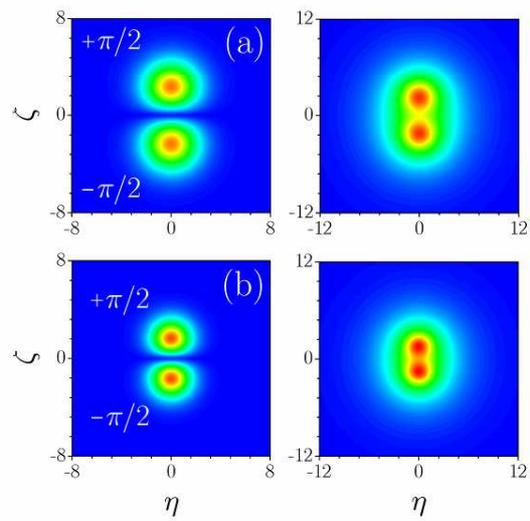

Figure 6 (color online). Field modulus (left column) and refractive index (right column) distributions for dipole solitons corresponding to (a) $b = 0.8$ and (b) $b = 1.8$ at $S = 0.06$, $d = 10$. All quantities are plotted in arbitrary dimensionless units.

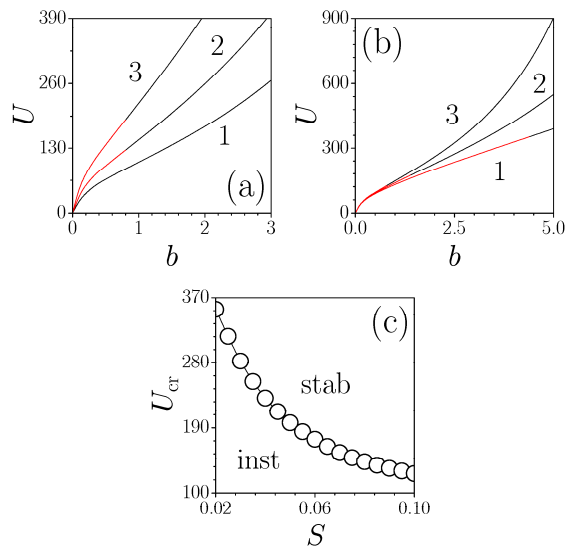

Figure 7 (color online). (a) Energy flow versus propagation constant for $d=5$ (curve 1), 10 (curve 2), and 20 (curve 3) at $S=0.1$. (b) Energy flow versus propagation constant for $S=0.02$ (curve 1), 0.06 (curve 2), and 0.1 (curve 3) at $d=10$. Black curves correspond to stable branches, red curves correspond to unstable branches. (c) Critical energy flow versus saturation parameter at $d=10$. All quantities are plotted in arbitrary dimensionless units.

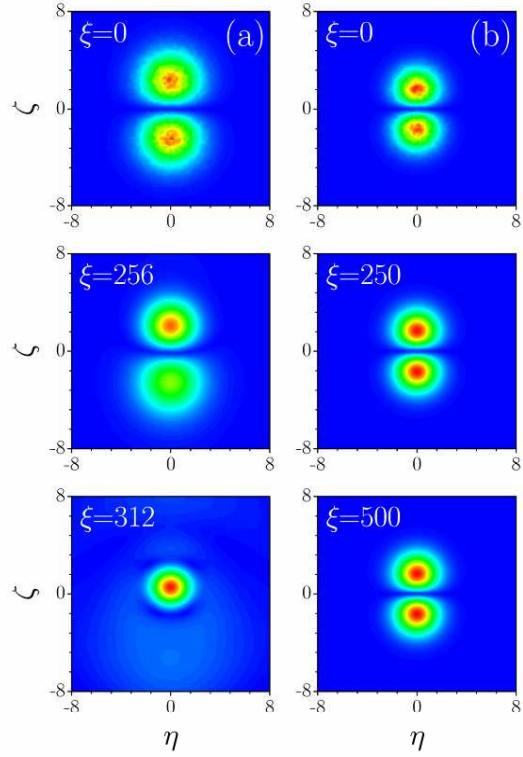

Figure 8 (color online). Propagation dynamics of (a) unstable dipole with $b = 0.8$ and (b) stable dipole with $b = 1.8$ in the presence of broadband input noise with variance $\sigma_{\text{noise}}^2 = 0.01$. In both cases $d = 10$, $S = 0.06$. Field modulus distributions are shown at different distances. All quantities are plotted in arbitrary dimensionless units.